\documentstyle[12pt,epsf]{article}

\begin{document}

\baselineskip 6mm
\renewcommand{\thefootnote}{\fnsymbol{footnote}}


\newcommand{\nc}{\newcommand}
\newcommand{\rnc}{\renewcommand}


\rnc{\baselinestretch}{1.24}    
\setlength{\jot}{6pt}       
\rnc{\arraystretch}{1.24}   

\makeatletter \rnc{\theequation}{\thesection.\arabic{equation}}
\@addtoreset{equation}{section} \makeatother



\def\be{\begin{equation}}
\def\ee{\end{equation}}
\def\ba{\begin{array}}
\def\ea{\end{array}}
\def\bea{\begin{eqnarray}}
\def\eea{\end{eqnarray}}
\def\nn{\nonumber\\}



\def\ct#1{\cite{#1}}
\def\la#1{\label{#1}}
\def\eq#1{(\ref{#1})}


\def\c#1{{\cal #1}}


\def\a{\alpha}
\def\b{\beta}
\def\g{\gamma}
\def\G{\Gamma}
\def\d{\delta}
\def\D{\Delta}
\def\ep{\epsilon}
\def\et{\eta}
\def\ph{\phi}
\def\Ph{\Phi}
\def\ps{\psi}
\def\Ps{\Psi}
\def\k{\kappa}
\def\l{\lambda}
\def\L{\Lambda}
\def\m{\mu}
\def\n{\nu}
\def\th{\theta}
\def\Th{\Theta}
\def\r{\rho}
\def\s{\sigma}
\def\S{\Sigma}
\def\ta{\tau}
\def\o{\omega}
\def\O{\Omega}

\def\pr{\prime}


\def\half{\frac{1}{2}}

\def\goto{\rightarrow}

\def\grad{\nabla}
\def\curl{\nabla\times}
\def\div{\nabla\cdot}
\def\pa{\partial}

\def\gl{\left\langle}
\def\gr{\right\rangle}
\def\bl{\left[}
\def\ml{\left\{}
\def\sl{\left(}
\def\el{\left.}
\def\er{\right.}
\def\br{\right]}
\def\mr{\right\}}
\def\sr{\right)}

\def\vac#1{\mid #1 \rangle}


\def\check{ \maltese {\bf Check!}}


\def\Tr{{\rm Tr}\,}
\def\det{{\rm det}}


\def\bc#1{\noindent {\bf $\bullet$ #1} \\ }
\def\ch {$<Check!>$ }
\def\ss {\vspace{1.5cm}}

\begin{titlepage}

\hfill\parbox{5cm} { }

\vspace{25mm}

\begin{center}
{\Large \bf Two Point Correlation Function
of Sine-Liouville Theory }

\vspace{15mm}
Jongwook Kim$^{\, a \,}$\footnote{jongwook@sogang.ac.kr},
Bum-Hoon Lee$^{\, a \,}$\footnote{bhl@ccs.sogang.ac.kr},
Chanyong Park$^{\, a \,}$\footnote{cyong21@ihanyang.ac.kr},
and Chaiho Rim$^{\, b \,}$\footnote{rim@chonbuk.ac.kr}
\\[10mm]

${}^a$ { Department of Physics, Sogang University,
Seoul 121-742, Korea} \\

${}^b$ { Department of Physics, Chonbuk National University,
Chonju 561-756, Korea}

\end{center}

\thispagestyle{empty}

\vskip2cm


\centerline{\bf ABSTRACT} \vskip 4mm

Exact two point correlation functions of sine-Liouville theory are
presented for primary fields with U(1) charge neutral, which may
either preserve or break winding number. Our result is checked
with perturbative calculation and is also consistent with previous
one which can be obtained by restricting the action parameters.

\noindent

PACS numbers: 11.25.Hf, 11.55.Ds

\vspace{2cm}


\end{titlepage}

\renewcommand{\thefootnote}{\arabic{footnote}}
\setcounter{footnote}{0}

\section{Introduction}
\noindent

Since early 80's the two-dimensional Liouville field theory (LFT) was recognized as the effective
field theory of the two-dimensional quantum gravity \ct{p}.
Moreover, there have been considerable
efforts to relate this area to the string theory, especially the non-critical string
theory \ct{ct,gn,hj}.

Recently, the interest in this LFT \ct{fzz,zz0}
was renewed with the observation of some dual relation
between these theories and the alternative matrix model \ct{kkm,bk,dj},
which was also described by the open string spectrum
on the D-branes. From the string point of view, this duality
could be interpreted as the open/closed string duality.

More interesting is
that there exists a 1+1-dimensional black hole solution in gravity side \ct{kkk}.
The string propagation in this geometry is described
by coset conformal field theory (CFT):
$SL(2,R)/U(1)$ for the Minkowski version and
$SL(2,C)/ \sl SU(2)\times U(1) \sr$ for the Euclidean one.
The question is how one can describe this black hole solution in terms of
matrix quantum mechanics (MQM) since it was noted
that there are no sufficient degrees of freedom for describing the black hole entropy
in the singlet part of MQM theory \ct{kms,fv}.

On the other hand, it was conjectured
that the string theory in the two dimensional black hole geometry
described by the SL(2,R)/U(1) coset conformal field theory
is dual to the (dual) sine-Liouville field theory (SLFT) with some special types
of the parameters.
This correspondence has not been proved but evidence for its validity has been
presented \ct{f,hk}.
It was also claimed in \ct{kkk,akk}
that the vortices having a winding number on the worldsheet correspond to
U(N) non-singlet states in MQM.
With the help of the connection between vortices and non-singlet states in MQM,
the integrable infinite Toda chain hierachy was constructed,
relating the usual c=1 string with SLFT backgrounds.
This allows one to compute the partition sum and the
correlation functions of string perturbation theory.
Thus, the study of vortices in the sine-Liouville theory
may disclose black hole information corresponding to entropy
and the Hawking temperature.

For integrable quantum field theory
defined as a perturbed CFT \ct{z},
LFT also proves an efficient tool to calculate
vacuum expectation values (VEV) of local fields.
In \ct{lz}, an explicit expression for the VEVs of the exponential fields
in the sine-Gordon and sinh-Gordon models was proposed
and in \ct{flzz} that this expression was shown to be obtained
as the minimal solution of certain reflection relations
which involve the Liouville reflection amplitude \ct{zz}.
The reflection amplitude idea was extended to boundary theories in \ct{bsG}.
In the presence of boundary the one-point function of LFT
is obtained in \ct{fzz,zz0}

In this paper, we will calculate the two-point correlation function
of bulk sine-Liouville theory.
In \ct{bf}, the two point function of exponential fields
was computed for the case when winding mode is preserved
and the action parameters
are restricted to satisfy a special relation.
We are considering the two-point function
with action parameters unrestricted.
In general, the two-point function may or may not preserve
the winding number.


The plan of the paper is as follows.
In section 2, we introduce the SLFT,
and its conformal algebra as well as degenerate operators.
In section 3, we use the degenerate operators
to find the recursion relation of two-point function
(the reflection amplitude) following Teschner \ct{t}.
The winding number violating two point function
is explicitly calculated and the result is checked
with perturbative calculation.
The result is also confirmed to coincide with the one in \ct{bf}
when the two point function is reduced to a special case
so that the the winding number is preserved.
section 4 is the conclusion and some remark is given.

\section{Sine-Liouville Theory and Corresponding Conformal Algebra}
Sine-Liouville field theory action is given in terms of three bosonic fields
\be
S_E = \frac{1}{16\pi} \int d^2 x \left\{
(\pa_{\m} \ph_1)^2 + (\pa_{\m} \ph_2)^2 + (\pa_{\m} \ph_3)^2
        - 32 \pi \m e^{\a\ph_1} \cos (\b \ph_2 + \g \ph_3)  + q \ph_1 {\c R}^{(2)} \right\} ,
\ee
where $q$ is a background charge
\be
q = \frac{1}{2\a} \left( 1 + \a^2 -\b^2 -\g^2 \right)
\ee
and ${\c R}^{(2)}$ is a two-dimensional curvature,
whose term makes this theory conformal.
For clarity, $\a$, $\b$  and $\g$ are considered real and positive.

Holomorphic energy-momentum tensor is given as
\be
T = - \frac{1}{4}  \left[  (\pa \ph_1)^2
 +  (\pa \ph_2)^2  +  (\pa \ph_3)^2  \right] + q \pa^2 \ph_1 \,,
\ee
whose composite operators are considered normal-ordered.
Using the quantum equation of motion, one can show the
conservation explicitly
\bea
\bar{\pa} T
&=& - 2\pi\m (1 - 2q \a) : \pa  e^{\a \ph} \cos (\b \ph_2 + \g \ph_3) :
\nn
 &&  -2\pi\m (\a^2 -\b^2 -\g^2) : \pa  e^{\a\ph} \cos (\b \ph_2 + \g \ph_3) :
\nn
 &= & 0 \,.
\la{conseq}
\eea
The first line  in the right hand side
comes from the classical equation of motion and the second line
from the quantum effect.

In the free field limit ($\ph_1 \to  -\infty$),
$U(1)\times U(1)$ symmetry is conserved;
\[
J_2 = i \pa \ph_2\,, \qquad J_3  = i \pa \ph_3 \,.
\]
In the sine-Liouville theory,
however, one of these U(1) symmetries is explicitly broken and
only one U(1) symmetry is preserved;
\be
J = i \left(\frac{1}{\b} \pa \ph_2 - \frac{1}{\g} \pa \ph_3 \right)\,.
\la{u1c}
\ee
The conserved current is given  under the transformation
$
\ph_2 \to \ph_2 + c \, \g , \; \ph_3 \to \ph_3 - c \, \b ,
$
with an arbitrary constant $c$,

Under the global transformation,
$
\b \pa \ph_2 + \g \pa \ph_3 \to \b \pa \ph_2 + \g \pa \ph_3 + 2 \pi N,
$
with an arbitrary integer number $N$,
the action is still invariant.  Therefore, it is useful to introduce
the broken $U(1)$ symmetry quantum number for later convenience.
The corresponding broken U(1) symmetry generator is defined as
\be
\tilde{J} (z) =  \frac{1}{(\b^2 +\g^2)} \left(\b \pa \ph_2 + \g \pa \ph_3 \right) \,,
\ee
and its quantum number is called soliton number or winding number.

The symmetry generators show the operator product expansion (OPE) as the following:
\bea
T(z) T(w) &=& \frac{c}{2(z-w)^4} + \frac{2}{(z-w)^2} T(w) + \frac{1}{(z-w)} \pa T(w)
                  + {\rm Reg} , \nn
J(z) J(w) &=& \frac{k}{(z-w)^2} +  {\rm Reg} , \nn
T(z) J(w) &=& \frac{1}{(z-w)^2} J(w) + \frac{1}{(z-w)} \pa J(w) + {\rm Reg}  , \la{ope}
\eea
where Reg means the regular term, $c$ is the central charge
\be
c = 3+24q^2 \,,
\ee
and  $k$ is the level of Kac-Moody algebra,
\be
k = 2 \left(\frac{1}{\b^2} + \frac{1}{\g^2} \right)
\ee
These symmetry generators introduce the Virasoro and the Kac-Moody algebra
\bea
\left[L_n , L_m \right] &=& (n-m) L_{n+m} + \frac{c}{12} (n^3 -n) \d_{n+m,0} \,, \nn
\left[J_n , J_m \right] &=& k n \d_{n+m,0} \,, \nn
\left[L_n , J_m \right] &=& -m J_{n+m} \,.
\eea

Normal-ordered vertex operator at a point $z$
is denoted as
\be
V(a,b,c;z) = : \exp [ a\ph_1 + i (b \ph_2 + c \ph_3)](z) : ,
\ee
where the  parameters $a,b$ and $c$ are arbitrary and real.
This primary operator has the OPE with $T(z)$ and $J(z)$,
\bea
T(z) V(a,b,c;w) &=& \frac{\D}{(z-w)^2} V(w) + \frac{1}{z-w} \pa V(w) + :T(w) V(w): \nn
                && + : \left( a \pa^2 \ph_1 (w) + ib \pa^2 \ph_2 (w) + ic \pa^2 \ph_3 (w) \right) V(w) :
                   + {\c O}(z-w),\nn
J(z) V(a,b,c;w) &=&  \frac{Q}{z-w} V(w) + : J(w) V(w) : + {\c O}(z-w) \,,
\eea
where $\D = \D (a,b,c) = a(2q-a) + b^2 +c^2$
is the conformal dimension
of the primary field $V(a,b,c; w)$,
and $Q=(b/\b -c/\g)$ is the U(1) charge.
From the broken generator $\tilde{J} (z)$,
we may define the winding number $\o = (\b b+ \g c)/ (\b^2 +\g^2)$.

The interaction term in the action
\be
V^{\pm} \equiv  \int d^2 x \,\, V(\a, \pm \b, \pm \g ;x) ,
\ee
is the screening operator,
whose conformal dimension and U(1) charge are roze.
Using this screening operator
one may construct degenerate operator with U(1) charge neutral \ct{df}.
The operator V(a,b,c; z) becomes degenerate when
\bea
a &=& - (m+n) \frac{\a_-}{2} - (\tilde{m}+\tilde{n}) \frac{\a_+}{2}  , \nn
b &=& - (m - n + \tilde{m} - \tilde{n}) \frac{\b}{2}  , \nn
c &=& - (m - n + \tilde{m} - \tilde{n}) \frac{\g}{2} ,
\eea
where
\be
\a_- = \a \quad {\rm and} \quad \a_+ = \frac{1-\b^2-\g^2}{\a} ~.
\ee
The degenerate operator with
$\tilde{m} = \tilde{n} = 0$ only will be used in our approach
since there is no interaction term of the type $e^{\a_+ \ph_1}$.
For notational simplicity, we will denote the degenerate operator as
\be
V_{(m,n)} (z) \equiv V(-(m+n) \frac{\a}{2}, -(m-n) \frac{\b}{2}, -(m-n) \frac{\g}{2}; z)\,,
\ee
and distinguish the degenerate ones
from other primary fields V(a,b,c; z) with $a,b,c$ arbitrary.

\section{Two-point correlation function}
\noindent

The two-point function
$\left< V(a,b,c; 1)\, V(a^{\pr},b^{\pr},c^{\pr}; 0) \right> $
is vanishing unless the two operators are of the same conformal dimension
and U(1) charge $Q$ is preserved.
We may normalize the two point function as
\be
\left< V(a,b,c; 1)\, V(2q-a,-b,-c; 0) \right> =1\,.
\ee
Thus one may consider a two-point correlation function
\be
R(a,b,c) \equiv \left< V(a,b,c; 1)\, V(a,-b,-c; 0) \right>   ,
\ee
which is called reflection amplitude.
Here the winding number is preserved.

A different reflection amplitude
\be
D(a,b,c) \equiv \left< V(a,b,c; 1) \,  V(a,b,c; 0) \right>
\ee
is also considered. The operators
have the right conformal dimension but violate the winding number.
Note that the action has the screening operator $V^{\pm}$,
changing the winding number by $\pm 1$ respectively.
Thus it is not surprising that
the winding number violating two-point function is non-vanishing.
Indeed, when $b/\b=c/\g$, the two-point function turns out to be non-vanishing.
This is the case $V{(a,b,c;z)}$ has the U(1) charge neutral $Q=0$.
If $\b=\g$,  even when $V{(a,b,c)}$ has  non-vanishing U(1) charge,
$D(a,b,c) $ may not vanish.

The reflection amplitudes in general satisfy the identity
\be
Z(a,b,c) = Z(a,-b,-c)  \quad {\rm and}  \quad Z(a,b,c)\,Z(2q-a,-b,-c)=1  , \la{symp}
\ee
where $Z$ can be either $R$ or $D$.
When $b=c=0$,  $D$ and $R$ coincide; $D(a,0,0) = R(a,0,0)$.

In our paper, we will concentrate on $D(a,b,c)$ function
with $U(1)$ neutral charge sector only.
For this purpose, we will
reserve the notation $D_0 (a,b,c)$ for this neutral sector,
instead of  $D (a,b,c)$.
To obtain $D_0(a,b,c)$ we may use the `Teschner' method \ct{t},
which uses the neutral degenerate operators to find
the functional relation between two-point functions.
For example, one may use
$V_{(1,0)}(z) = V(-\frac{\a}{2}, -\frac{\b}{2}, -\frac{\g}{2}; z)$
to obtain
\be
C_{(1,0)}(a,b,c) = \frac{D_0 (a,b,c)}{D_0 (a+\frac{\a}{2},b+\frac{\b}{2},c+\frac{\g}{2})} ,
\ee
where $C_{(1,0)}(a,b,c)$ is the structure constant in the OPE
\be
V_{(1,0)} \bigotimes V(a,b,c) = V(a - \a/2, b - \b/2, c - \g/2) + C_{(1,0)} (a,b,c)
             V(a + \a/2, b + \b/2, c + \g/2) .
\ee

In the sine-Liouville theory, however, consistency problem arises
at level two. Note that at level two,
two degenerate operators exist,
$V_{(1,0)}$ and $V_{(1,0)}$.
Let us consider  the OPE
\bea
V_{(1,0)} \bigotimes V_{(0,1)} &=& V_{(1,1)} + V_{(1,-1)} \nn
                               &=& V_{(1,1)} + V_{(-1,1)} .
\eea
When $V_{(1,0)}$ and $V_{(0,1)}$ are used simultaneously,
we have to get rid of the effect, $V_{(1,-1)}$ or $V_{(-1,1)}$
to obtain the restricted Hilbert space.
The problem is that functional approach
does not realize this restriction easily,
using the screening operator insertions.
Thus, it is safe to obtain the functional relation
$V_{(1,1)}$, instead of using $V_{(1,0)}$ and $V_{(0,1)}$ simultaneously.
The same thing applies for $V_{(m,n)}$, in general.

\subsection{The simplest degenerate operator case, $m+n=1$}
\noindent

We will consider the simplest case first,
$(m,n)=(1,0)$ or $(m,n)=(0,1)$.
The fusion rule of the neutral vertex operator with  $V_{(1,0)}$
($V_{(0,1)}$) is given as
\bea
&&V(a,b,c) \bigotimes V_{(1,0)} \nn
 && \quad  =  C_0 \,\, V(a - \a/2, b - \b/2, c - \g/2) + C_{(1,0)} \,\,
             V(a + \a/2, b + \b/2, c + \g/2) , \label{fusion}
\eea
or
\bea
&&V(a,b,c) \bigotimes  V_{(0,1)} \nn
 && \quad =  C_0 \,\, V(a - \a/2, b + \b/2, c + \g/2)
     + C_{(0,1)} \,\,
             V(a + \a/2, b - \b/2, c - \g/2) , \label{fusion}
\eea
where $C_{(1,0)} $ and $C_{(0,1)} $ are the structure constants.
Here, we set $C_0 =1$ since we do not have to insert screening operators.
Then, $C_{(1,0)} $ and $C_{(0,1)} $ are given as
\be
C_{(1,0)} =C_{(1,0)} (a,b,c) = - \m \pi \frac{\g(-1 + 2\a a - 2\b b - 2\g c -\a^2 + \b^2 + \g^2 )}
                  {\g(-\a^2+\b^2+\g^2) \g(2\a a - 2\b b - 2\g c)} , \la{defc-}
\ee
and
\be
C_{(0,1)} =C_{(0,1)} (a,b,c) = - \m \pi \frac{\g(-1 + 2\a a + 2\b b + 2\g c -\a^2 + \b^2 + \g^2 )}
                  {\g(-\a^2+\b^2+\g^2) \g(2\a a + 2\b b + 2\g c)}\,,  \la{defc-}
\ee
where $\g(x)= \G(x) / \G(1-x)$.

Using the associativity  of the three-point function,
the functional relation between $D_0$'s is given as
\bea
&& \left< V(a,b,c; 0)\,\, V_{(1,0)} (1)\,\,
           V(a + \frac{\a}{2}, b + \frac{\b}{2}, c + \frac{\g}{2} ;\infty) \right> \nn
   \quad \quad &=& D_0 (a,b,c)
= C_{(1,0)} (a,b,c)\,\,
D_0 (a + \frac{\a}{2}, b + \frac{\b}{2}, c + \frac{\g}{2}) \,,
\la{shift}
\eea
which holds when $a$, $b$ and $c$ satisfy
$ (a-q)/{\a} = - b/{\b}= - c/{\g} $.
The exact two point function is given by
\be
D_0 (a,b,c) = \left[ - \pi \m \g(\a^2 - \b^2 -\g^2) \right]^{(2q-2a)/\a}
           \frac{\g(2\a a - 2\b b - 2\g c - \a^2 + \b^2 + \g^2)}
           {\g \left( 2 - \frac{2\a a - 2\b b - 2\g c}{\a^2 - \b^2 - \g^2}
           + \frac{1}{\a^2 - \b^2 - \g^2} \right)} , \la{ream-}
\ee
This two-point function satisfies the symmetric properties given in \eq{symp}.

When  $ (a-q)/{\a} =  b/{\b} =  c/{\g}$,
we have to use $V_{(0,1)}$ instead of
$V_{(1,0)}$.
In this case, we have the similar relation with \eq{shift} where $\b$ and $\g$ are
replaced with $-\b$ and $-\g$;
\be
D_0(a,b,c) = \left[ - \pi \m \g(\a^2 - \b^2 -\g^2) \right]^{(2q-2a)/\a}
           \frac{\g(2\a a + 2\b b + 2\g c - \a^2 + \b^2 + \g^2)}
           {\g \left( 2 - \frac{2\a a + 2\b b + 2\g c}{\a^2 - \b^2 - \g^2}
           + \frac{1}{\a^2 - \b^2 - \g^2}\right)} . \la{ream+}
\ee


One might check the correctness of this result using perturbative calculation
when $a<q$.
Suppose
\be
a =  q - n \frac{\a}{2}  , \quad b = - n \frac{\b}{2}  \quad {\rm and} \quad
c = - n \frac{\g}{2}       \la{cons1}
\ee
with a positive integer $n$.
With $n$ screening operators we have
\bea
D_0 (a,b,c) &=& \frac{(-\mu)^n}{n!} \prod_{i=1}^{n} \int d^2 z_i
            \left<  V(a,b,c;0) \,\, V(a,b,c;\infty) \,\, V(\a,\b,\g;z_i) \right>  \nn
         &=& \frac{(-\mu)^n}{n!} \prod_{i=1}^{n} \int d^2 z_i |z_i|^{-4 (\a a - \b b - \g c)}
             \prod_{i<j}^{n} |z_i - z_j|^{-4(\a^2 -\b^2 -\g^2)} . \la{nref}
\eea
Note that this integration does not converge due to the zero-mode integration
and needs regularization.
For this purpose,
we slightly modify $a$ into $2\a a = -n \a^2 + 2q\a + \ep$ where $\ep$ is small. Then, \eq{nref} is rewritten as
\bea
D_0 (a,b,c) &=& \frac{(-\m)^n}{n} \int d^2 x_1 |x_1|^{-2-2n\ep} \nn
          && \times \frac{1}{(n-1)!}\prod_{i=2}^n \int d^2 y_i |y_i|^{-4(\a a-\b b-\g c)}
              |1-y_i|^{-4(\a^2-\b^2-\g^2)} \nn
          && \times \prod_{i=2, i<j} |y_i -y_j|^{-4(\a^2-\b^2-\g^2)} . \la{int1}
\eea
The first term is still divergent at $|x_1| = 0$ when $\ep > 0$.
We further introduce a UV cut-off $\L$
\be
\int_{\L}^{\infty} d^2 x_1 |x_1|^{-2-2n\ep} =  \frac{\pi}{n\ep} \L^{-2n\ep} .
\ee
Redefining the renormalized parameter $\tilde{\m} = \m \L^{-2\ep}$ and performing the rest of the integral in
\eq{int1}, we finally obtain
\be
D_0 (a,b,c) = \frac{(-\pi \tilde{\m})^n}{(n!)^2} \bl \g(\a^2 -\b^2-\g^2) \br^n
           \g \sl (1-n(\a^2 -\b^2-\g^2) \sr \frac{1}{\ep} . \la{perres}
\ee
This perturbative result
coincides with the proposed exact correlation function
in \eq{ream-} (\eq{ream+})
if the vertex operator parameter has the value in \eq{cons1}
and the cosmological constant is identified with the
renormalized one $\tilde{\m}$.


This simplest case result with $m+n=1$ can be extended to the case
$(m,n)=(m,0)$ or $(m,n)=(0,n)$ easily.
Noting that the structure constant $C_{(m,0)}$ in
$V_{(m,0)} \bigotimes V(a,b,c)$ is given by a product form of $C_{(1,0)}$'s.
\be
C_{(m,0)} (a,b,c) = \prod_{p=0}^{m-1} C_{(1,0)} \sl a+p\frac{\a}{2},
b+p\frac{\b}{2},
                    c+p\frac{\g}{2} \sr .
\ee
One has the functional relation
\bea
&& C_{(m,0)} (a,b,c) \nn
&& \quad =  \left< V(a,b,c;0) V (-m \frac{\a}{2}, -m \frac{\b}{2}, -m \frac{\g}{2};1)
         V(2q -a -\frac{\a}{2}, - b -m \frac{\b}{2}, - c -m \frac{\g}{2} ;\infty ) \right> \nn
&& \quad = \frac{D_0 (a,b,c)}{D_0 (a+\frac{m}{2}\a, b+\frac{m}{2}\b , c+\frac{m}{2}\g)} ,
\eea
which is consistent with the result in \eq{shift}.
Likewise, the similar relation holds when $(0,n)$ if $(\b,\g) \to (-\b,-\g)$.

\subsection{Case with $m+n=2$}

When $(m,n)=(2,0)$ or $(m,n)=(0,2)$,  the result is obtained in the previous section.
When $(m,n)=(1,1)$, the degenerate operator has vanishing winding number and
can be used to obtain the winding number preserving two-point function. Note that
$D_0 (a,0,0) =R(a,0,0)$ and $D_0 (a,0,0)$ preserves the winding number in this case.


From the fusion rule
\be
V_{(2,0)} \bigotimes V(a,0,0)
= V(a-\a,0,0) +   C_{(1,1)}\,\,  V(a+\a,0,0) ,\la{deg2}
\ee
the fusion coefficient $C_{(1,1)} $  is given as
\be
C_{(1,1)} =C_{(1,1)} (a,0,0)
=  \left< V(a,0,0)\, V_{(1,1)} \, V(2q-(a+\a),0,0) \right> \,.
\ee
This can be calculated using one screening operator  ($V^+ \,V^-$).
\be
C_{(1,1)} (a,0,0)  =  \xi \times \frac{\g(-1 + 2\a a - \a^2 +\b^2 +\g^2) \g( -1 + 2\a a +2\b^2 +2\g^2)}
               {\g(2\a a) \g(2\a a + \a^2 +\b^2 +\g^2)}  , \la{coe1}
\ee
where
\be
\xi = (-\pi \m)^2 \frac{\g(-2 \sl \a^2 + \b^2 + \g^2) \sr}{\g \sl -(\a^2 + \b^2 + \g^2)\sr} \g(1+2\a^2)
            \g(1+\a^2 -\b^2 -\g^2) .  \la{annor}
\ee
The recursion relation between $D_0 (a,0,0)$ and $D_0 (a+\a,0,0)$ is given as
\bea
\left< V(a,0,0) V_{(1,1)} V(a+\a,0,0) \right> &=& C_{(1,1)}(a,0,0) D_0 (a+\a,0,0) \nn
                                         &=&  D_0 (a,0,0) \la{recrel2} .
\eea
By applying \eq{recrel2} recursively, we  obtain $D(a+n\a,0,0)$
\be
D_0 ( a+n\a,0,0) = \prod_{p=0}^{n-1}
\frac{1}{ C_{(1,1)} (a+p\a,0,0)} D_0 (a,0,0) . \la{rerelf}
\ee
To find the closed form, we use the integral representation of the $\G$-function
\be
\log \G(z) = \int_0^{\infty} \frac{dt}{t} \ml \frac{e^{-zt} -e^{-t}}{1-e^{-t}} + (z-1) e^{-t} \mr ,
\ee
and start from $a=q$. With $D_0(q,0,0)$=1, $D_0(q+n\a,0,0)$ is rewritten as
\bea
&& D_0 (a,0,0) \nn
&& =  \sl \xi \sr^{\frac{q-a}{\a}}
             \frac{\G \sl 1+ \frac{a-q}{\a} \sr        \G \sl \frac{q}{\a} -\frac{a-q}{\a} \sr
                   \G \sl 1+2\a (a - q)  \sr       \G \sl 1-2\a (a-q) +2\a q   \sr}
                  {\G \sl 1- \frac{a-q}{\a}   \sr        \G \sl \frac{q}{\a} + \frac{a-q}{\a}    \sr
                   \G \sl 1-2\a (a - q)   \sr\           \G \sl  1+2\a (a-q) +2\a q  \sr } \nn
&& \times \exp \bl \int_0^{\infty} \frac{dt}{t} \ml \frac{e^{-(1+2\a^2+2\a a-4\a q)t}
            {- e^{-(1+2\a a +4\a q)t}}}{(1-e^{-t}) (1-e^{-2\a^2 t})}  \sl 1- e^{-4\a qt} \sr
             - 8(a-q) q e^{-t}  \mr \br  \la{gesol},
\eea
where $a=q+n\a$. Even though $D_0 (a,0,0)$ is obtained when $a$ is a discrete number,
we can analytically continue to arbitrary $a$.

Suppose $4\a q = m$ with $m$ integer, then $D_0(a,0,0)$ is explicitly given
in terms of gamma function:
\bea
&& D_0(a,0,0)  \nn
&& \quad =  \ml \sl \frac{1}{2\a^2} \sr^{8\a q} \xi \mr^{\frac{q-a}{\a}}
             \frac{\G \sl 1+ \frac{a-q}{\a} \sr        \G \sl \frac{q}{\a} -\frac{a-q}{\a} \sr
                   \G \sl 1+2\a (a - q)  \sr       \G \sl 1-2\a (a-q) +2\a q   \sr}
                  {\G \sl 1- \frac{a-q}{\a}   \sr        \G \sl \frac{q}{\a} + \frac{a-q}{\a}    \sr
                   \G \sl 1-2\a (a - q)   \sr\           \G \sl  1+2\a (a-q) +2\a q  \sr } \nn
&& \quad \times \prod_{p=0}^{m-1} \frac{\G \sl 1 + \frac{1+2\a a - 4\a q + p}{2\a^2} \sr}
                                {\G \sl 1 + \frac{1-2\a a+ p}{2\a^2}\sr} . \la{sol1}
\eea
For the special case $m=1$ ($2\a q = \half$ or $\a^2 - \b^2 -\g^2 = -\half$),
$D_0(a,0,0)$ is given by
\bea
D(a,0,0) &=&  \sl \frac{(\pi \m)^2}{4\a^4} 2^{-8\a^2-3} \sr^{\frac{q-a}{\a}} \nn
         &&  \times  \frac{\G \sl 1+2\a(a-q) \sr        \G \sl \half - 2\a(a-q) \sr
                   \G \sl \frac{a-q}{\a} \sr }  {    \G \sl 1 - 2\a(a-q) \sr
                  \G \sl \half + 2\a(a-q)  \sr        \G \sl - \frac{a-q}{\a}   \sr } ,
\eea
which reduces to the  exactly same result given in  \ct{bf} for $b=c=0$.

When $4\a q = 2\a^2 m$ with integer $m$,  $D_0(a,0,0)$ is given as
\bea
&& D(a,0,0)  \nn
&& \quad = \sl \xi \sr^{\frac{q-a}{\a}}
            \frac{\G \sl 1+ \frac{a-q}{\a} \sr        \G \sl \frac{q}{\a} -\frac{a-q}{\a} \sr
                   \G \sl 1+2\a (a - q)  \sr       \G \sl 1-2\a (a-q) +2\a q   \sr}
                  {\G \sl 1- \frac{a-q}{\a}   \sr        \G \sl \frac{q}{\a} + \frac{a-q}{\a}    \sr
                   \G \sl 1-2\a (a - q)   \sr\           \G \sl  1+2\a (a-q) +2\a q  \sr } \nn
&& \quad \times \prod_{p=0}^{m-1} \frac{\G \sl 1 + 2(p+1) \a^2 + 2\a (a-q) - 2\a q \sr}
                                {\G \sl 1 + 2 (p+1) \a^2 - 2\a (a-q) -2\a q \sr} .
\eea

\section{Conclusion}

\indent

In this paper, we present two types of two-point correlation
functions (reflection amplitudes) of the sine-Liouville field theory:
One preserves the winding number and the other is not.
This winding number violating process might be useful to understand
the U(N) non-singlet part in MQM, similar
to the spectral flow giving stringy effects \ct{mo}.
This is the difference of sine-Liouville theory
from the usual Liouville or N=1 super Liouville field theory
\ct{n=1} case, which does not support winding number process.

It is also noted that
due to the existence of neutral degenerate operators at the same level
with opposite non-vanishing winding number,
the Hilbert space cannot be reduced easily in functional formalism.
Without proper treatment of the Hilbert space,
one may lead to inconsistency in
the functional relations of correlation functions.
To overcome this difficulty, we avoid using
the product of degenerate operators to find the functional relations
of two-point functions.

The explicit result of $D(a,b,c)$ in the $U(1)$ neutral sector
is given for the cases: $(a-q)/ \a = - b/ \b= - c/\g$, $(a-q)/ \a =  b/ \b=  c/\g$,
and $b=c=0$ with $a$ arbitrary.
One may elaborate the calculation further.
Suppose $(m,n)$ is co-prime and
the primary field is given as $(a-q)/b = t  \a/ \b $
with $t =(m+n)/(m-n)$ ($|t| >1$).
In this case, the functional relation is given as
\be
D_0(a,b,c) = C_{(m,n)}(a,b,c) \;\; D_0(a+(m+n)\frac{\a}{2},b+(m-n)\frac{\b}{2},c+(m-n)\frac{\g}{2}),
\ee
and
\bea
C_{(m,n)}(a,b,c) &=& \frac{(-\m)^{m+n}}{(m+n)!} \prod_{i=1}^{m} \int d^2 x_i |x_i|^{-4(\a a-\b b-\g c)}
                     |1-x_i|^{2 \ml (m+n)\a^2-(m-n)\b^2 -(m-n)\g^2 \mr} \nn
                 && \quad \times \prod_{i^{\pr}=1}^{n} \int d^2 y_{i^{\pr}} |y_{i^{\pr}}|^{-4(\a a+\b b+\g c)}
                     |1-y_{i^{\pr}}|^{2 \ml (m+n)\a^2+(m-n)\b^2 +(m-n)\g^2 \mr} \nn
                 && \quad \times \prod_{i<j}^{m} |x_i-x_j|^{-4(\a^2-\b^2-\g^2)} \;\;
                           \prod_{i^{\pr}<j^{\pr}}^{n} |y_{i^{\pr}}-y_{j^{\pr}}|^{-4(\a^2-\b^2-\g^2)} \nn
                 && \quad \times \prod_{(i,i^{\pr})}^{(m,n)} |x_i-y_{i^{\pr}}|^{-4(\a^2+\b^2+\g^2)} \;\;\; .
\eea

Unfortunately, this integration is not given in closed form yet,
even though similar integration is present in \ct{fh}.
The explicit form of the reflection amplitude with arbitray vertex operators
might be desirable in a near future,
considering the relation of the sine-Liouville theory
with black hole entropy
and with N=2 super-Liouville theory
\ct{h, n=2,eguchi}
which may be viewed as a special case of the sine-Liouville theory \ct{f}.

\vspace{1cm}

\noindent{\bf Acknowledgement}

This work was supported in part
by the Korean Research Foundation Grant KRF 2003-015-C00111 (B. H. Lee),
KRF 2003-070-C00011 (J. Kim and C. Park),
the Research Institute for Basic Science of Sogang University (C. Park),
the Basic Research Program of the Korea Science
and Engineering Foundation Grant number  R01-2004-000-10526-0 (C. Rim).

\vspace{1cm}

\nc{\np}[3]{Nucl. Phys. {\bf B#1}, #2 (#3)}

\nc{\plb}[3]{Phys. Lett. {\bf B#1}, #2 (#3)}

\nc{\prl}[3]{Phys. Rev. Lett. {\bf #1}, #2 (#3)}

\nc{\prd}[3]{Phys. Rev. {\bf D#1}, #2 (#3)}

\nc{\ap}[3]{Ann. Phys. {\bf #1}, #2 (#3)}

\nc{\prep}[3]{Phys. Rep. {\bf #1}, #2 (#3)}

\nc{\ptp}[3]{Prog. Theor. Phys. {\bf #1}, #2 (#3)}

\nc{\rmp}[3]{Rev. Mod. Phys. {\bf #1}, #2 (#3)}

\nc{\cmp}[3]{Comm. Math. Phys. {\bf #1}, #2 (#3)}

\nc{\mpl}[3]{Mod. Phys. Lett. {\bf A#1}, #2 (#3)}

\nc{\cqg}[3]{Class. Quant. Grav. {\bf #1}, #2 (#3)}

\nc{\jhep}[3]{J. High Energy Phys. {\bf #1}, #2 (#3)}

\nc{\hep}[1]{{\tt hep-th/{#1}}}

\nc{\app}[3]{Ann. Phys. {\bf #1}, #2, (#3)}

\nc{\prp}[3]{Phys. Rept. {\bf #1}, #2, (#3)}

\nc{\jmp}[3]{J. Math. Phys. {\bf #1}, #2, (#3)}


\end{document}